\documentclass[5p,authoryear]{elsarticle}

\usepackage{graphicx}

\usepackage{amssymb}

\begin{document}

\title{PSDF: Particle Stream Data Format for N-Body Simulations}

\author[WMF]{Will M. Farr\corref{cor1}}
\ead{w-farr@northwestern.edu}
\author[PH]{Jeff Ames}
\author[PH]{Piet Hut}
\author[JM]{Junichiro Makino}
\author[SM]{Steve McMillan}
\author[TM]{Takayuki Muranushi}
\author[KN]{Koichi Nakamura}
\author[KN2]{Keigo Nitadori}
\author[SPZ]{Simon Portegies Zwart}

\cortext[cor1]{Corresponding author}

\address[WMF]{Northwestern University Center for Interdisciplinary
  Research in Astrophysics, 2145 Sheridan Rd., Evanston IL 60208 USA}

\address[PH]{Institute for Advanced Study, Princeton, NJ 08540, USA}

\address[JM]{Interactive Research Center of Science, Graduate
  School of Science and Engineering Tokyo Institute of Technology,
  2--12--1 Ookayama, Meguro, Tokyo 152-8551, Japan}

\address[SM]{Department of Physics, Drexel University, 3141 Chestnut
  Street, Philadelphia, PA 19104}

\address[TM]{The Hakubi Center, Kyoto University, Kitashirakawa Oiwakecho,
Sakyo-Ku, Kyoto, 606-8502, Japan}

\address[KN]{Graduate School of Information Science and Technology,
  The University of Tokyo.  7-3-1 Hongo, Bunkyo-ku, Tokyo, Japan}

\address[KN2]{Center for Computational Science, University of Tsukuba,
  1–1–1, Tennodai, Tsukuba, Ibaraki 305–8577, Japan}

\address[SPZ]{Leiden University, P.O. Box 9513, 2300 RA Leiden, The
  Netherlands}

\begin{abstract}
  We present a data format for the output of general N-body
  simulations, allowing the presence of individual time steps.  By
  specifying a standard, different N-body integrators and different
  visualization and analysis programs can all share the simulation
  data, independent of the type of programs used to produce the data.
  Our Particle Stream Data Format, PSDF, is specified in YAML, based
  on the same approach as XML but with a simpler syntax.  Together
  with a specification of PSDF, we provide background and motivation,
  as well as specific examples in a variety of computer languages.  We
  also offer a web site from which these examples can be retrieved, in
  order to make it easy to augment existing codes in order to give
  them the option to produce PSDF output.
\end{abstract}


\maketitle

\section{Introduction}

The simplest N-body calculations use a shared time step length for all
particles, implying a straightforward structure of the output.  With
N particles and k time steps, the output takes on the form of an $N*k$
matrix of particle data, where the latter typically contain the mass,
position and velocity of a single particle at a specific time, with
possible additional information such as higher derivatives of the
position (acceleration, jerk, etc.), the value of the potential at the
position of that particle, and so on.  The output of this matrix can
be done by ordering in time or by ordering by the identity of
particles, in which case each particle's history is output separately.

Some complications may occur when particles are removed, for example
because they are escaping from the system, or because they represent a
star that undergoes a destructive supernova leaving no remnant.
However, the basic I/O structure is simple enough that it is easy to
present these kinds of data in one of the standard data formats, such
as FITS \citep{Pence2010} or HDF \citep{HDF2011}, with a brief
description.

The situation becomes vastly more complicated when we allow for
individual time steps.  Simulations of dense stellar systems, such as
open and globular star clusters, as well as galactic nuclei, have
relied on the use of individual time steps, at least since the 1960s
\citep{Aarseth2009}.  The reason is that the presence of close
binaries and triples in such systems would increase the computer power
needed by orders of magnitude in case of shared time steps, compared
to individual time steps.  In addition, cosmological codes, too, often
use individual timesteps, given the increasingly large discrepancies
of intrinsic time scales that come with increasingly high spatial
resolution \citep[e.g.][]{Springel2005}.

The simplest way to output data from individual time step codes would
be to use shared time steps for the output.  Indeed, typical legacy
codes, such as NBODY6, do just that by default.  If all one wants to
do is to make a fixed movie of a simulation run, that approach
suffices.  However, when we interactively inspect the results of a
simulation run, we want to be able to zoom in and out, and speed up
and slow down the rate at which we run the graphics presentation of
the run.  With a fixed initial output rate, it may not be possible to
interpolate the motion of the particles that move at high speeds.
Phrased differently, an output rate high enough to faithfully present
the motion of all particles may be prohibitively expensive in terms of
memory.  It would be much better to let the graphics program itself
decide how and where to extrapolate, given the original data it has
received from a simulations code.

For example, when we display the dense center of a star cluster, the
graphics program can then use the full information for the rapidly
moving particles, while interpolating the data for the slower halo
particles.  Such an approach can easily save orders of magnitude of
memory storage requirement.  This approach was implemented by McMillan
within the ``Starlab'' simulation environment
\citep{PortegiesZwart2001}.  The associated {\tt tdyn} data format is
described in Section 10 of \citet{Hut2001} and explored in more detail
by \citet{Faber2010}.  However, this implementation was handcrafted
for a specific code, reflecting the data structure used in that code.
Clearly, it would be desirable to have a more universal data format
that allows different codes to share data in a more transparent way.

Other concerns are to make a data format standard machine independent,
to make allowance for parallel processing, and to avoid serious
overhead penalties with respect to performance.

Here we describe the ``Particle Stream Data Format,'' or PSDF, a
machine-independent, algorithm-independent data format for storing the
results of a simulation of point-mass gravitational dynamics using
individual timesteps.

\section{Basic idea}

We wish to store the evolution of a gravitating system of $N$ bodies
throughout a simulation with individual timesteps for each
particle. Conceptually, what we need is a stream of phase-space
information of particles, updated each time the integration algorithm
adjusts a particle's phase space information.  One possibility for
such a stream could be:
\begin{verbatim}
  particle_id, time, mass, x, y, z, vx, vy, vz, ...
  particle_id, time, mass, x, y, z, vx, vy, vz, ...
  particle_id, time, mass, x, y, z, vx, vy, vz, ...
\end{verbatim}
However, the data format should be flexible enough to be able to
include more information, if available, such as
\begin{itemize}
\item hierarchical decompositions of the system into binaries,
  triples, etc.
\item radius, and other info related to stellar evolution
\item merger history
\item fluid properties if a particle is an SPH particle
\item close encounter history
\item stellar evolution history
\item and so on.
\end{itemize}

One way to construct such a flexible data format is to use a 
self-describing data format, such as XML or YAML. For simplicity, we
adopt YAML \citep{YAML2011} here; there are libraries for reading and
writing YAML in many popular programming languages, and the format is
simple enough to be understood easily by humans, even if they are not
already familiar with it.

\subsection{Some basics of YAML}

The following is a simple example of data in YAML format.
\begin{verbatim}
--- !Particle
id: 0
r:
  - 0.1
  - 0.2
  - 0.3
v:
  - -1
  - -2
  - -3
m: 1.0
\end{verbatim}
In the above example, the line
\begin{verbatim}
--- !Particle
\end{verbatim}
Is the header, which indicates that it describes the data of an object
of type {\tt Particle}.  The line
\begin{verbatim}
id: 0
\end{verbatim}
defines a field with name ``id'', and value 0.  The text
\begin{verbatim}
r:
  - 0.1
  - 0.2
  - 0.3
\end{verbatim}
means the field ``r'' is an array with three elements. The first ``-''
means this line is a data for an array.  By default, numbers without
``.'' are regarded as integers, and with ``.'' floating point. Note
that indentation has meaning here and ``-'' must be indented the same
level or deeper than ``r'' and should be aligned.  The ``v'' and ``m''
fields behave similarly.  The order of the fields is not important.
The {\tt Particle} object behaves as a \emph{map} from names, like
``r'', to values, in the case above the array ``\verb|[0.1, 0.2, 0.3]|''.

\section{Particle Stream Data Format}

With the minimal description of YAML in the previous section, we can
now define the Particle Stream Data Format: the data format is a
stream of YAML representations of particle objects.  For example, a
valid PSDF fragment is
\begin{verbatim}
--- !Particle
id: 0
t: 0
r:
  - 0.1
  - 0.2
  - 0.3
v:
  - -1
  - -2
  - -3
m: 1.0
--- !Particle
id: 1
t: 0
r:
  - 0.2
  - 0.3
  - 0.4
v:
  - 0
  - 0
  - 0
m: 1.0
\end{verbatim}
This fragment describes two particles, at $t = 0$, with ids 0 and 1.

Particle objects behave in YAML as mappings from names to values.
Therefore, a specification of the meaning of certain names and a
procedure for handling unknown names in the stream are sufficient to
define the data format.  In Table \ref{tab:names} we list the reserved
names of our PSDF.  Any particular particle in a PSDF stream need not
include a value in its map for any of these names, but if it does, the
value must have the meaning in the table; similarly, if a particle
object in a PSDF stream does contain a value with one of the meanings
in the table, then it should be identified by the corresponding name.
Note that these requirements allow for easy extension of a PSDF stream
with application-specific information by including any names and
values not in Table \ref{tab:names} needed by the specialized
application.  Programs that understand this additional information can
benefit, while those that do not will still be able to function using
the basic information from any included values of Table
\ref{tab:names}.  In the future, we intend to provide extensions that
are useful for hierarchical decomposition of an $N$-body system, for
example one containing tight binaries, triples, and higher multiples,
and for description of fluid SPH particles.

\begin{table}
\begin{tabular}{|c|l|}
\hline
Name & Meaning\\
\hline
id & index (can be arbitrary text)\\
m & mass\\
t & time\\
t\_max & max time to which this record is valid \\
r  & position, array with three elements\\
v  & velocity, array with three elements\\
pot  & potential\\
acc  & acceleration, array with three elements\\
jerk  & jerk, array with three elements\\
snap  & snap, array with three elements\\
crackle  & crackle, array with three elements\\
pop  & pop, array with three elements\\
\hline
\end{tabular}
\caption{\label{tab:names} The reserved names of PSDF and their
  meanings.  Other names occurring in the data stream are to be
  ignored if they are not meaningful to an application or interpreted
  in an application-specific way if they are.  Here the jerk,
  snap, cracle, and pop are our names for the third, fourth, fifth,
  and sixth derivatives of position.  We specify that each vector must
  contain three elements; in the event of a two-dimensional
  simulation, one of the vector components should be set to zero.}
\end{table}

We require that time, position, velocity and higher derivatives are
consistent (for example, if position is given in parsecs and time in
years, velocity must be in parsec/year).  The name {\tt t\_max} is
rather special, in that it does not specify part of the state of a
particle, but rather gives the maximum possible time that this record
is used to predict the orbit of this particle; we expect that this may
prove useful to prevent invalid extrapolation in programs that process
the PSDF.  If no {\tt t\_max} is given, it should be assumed that the
particle record is valid until an updated record is encountered in the
stream, or forever if no such updated record appears subsequently in
the stream.



A complete PSDF object is a stream of particle objects describing the
states of individual particles in the system at particular times.
Such a stream may be consumed as it is produced, as in the case of an
integrator program whose output is directed to a graphics program that
displays the result of the integration; or such a stream may we
written to one or more files to be processed at a later date.  We do
not impose any particular ordering on the particle records in a PSDF
stream.  For some applications an ordering in time may be appropriate,
while for others an ordering in particle id may be better, or even
more complex orderings; in Section \ref{sec:repository} we provide
references to code that can convert between the time and particle-id
orderings.

\section{Rationale}

Our goal is to describe a data format that is 
\begin{enumerate}
\item Space-efficient for storing the data from individual-timestep
  $N$-body simulations.
\item Simple enough to be human readable and writeable, and safe for
  programs to read, even from un-trusted sources.
\item Information-rich to allow for post-processing and analysis or
  even provide enough additional information for continuation or
  re-running of a simulation.
\item Flexible enough to accommodate the special needs of programs
  that have more complex objects than point-mass particles.
\item Composable, so that fragments of the data can be split off for
  separate analysis and recombined easily.
\end{enumerate}
In this section we describe how these design goals led to the
specification in the previous section.

\subsection{Space Efficient}

As outlined in the introduction, it is extremely wasteful to produce a
complete system snapshot in an $N$-body simulation every time some of
the bodies update their positions or velocities.  The PSDF format
allows for the output of only the changed data---the new states of the
updated particles at the new time---to the stream.  Though the native
format is text-based, for readability, we have found that common
compression algorithms such as {\tt gzip} applied to files containing
PSDF data produce output that is within 10\% of the size of equivalent
compressed binary data.

\subsection{Simple and Human-Readable}

By using a general format (YAML) in wide use for our PSDF, we ensure
that there are mature, debugged libraries available for reading and
writing our format \citep{YAML2011}.  However, YAML is simple enough
that it can we written and modified easily ``by hand'' in a text
editor using only ASCII characters.  The idea of streaming updates to
individual particle states also meshes nicely with the evolution
algorithms in most $N$-body integrators, making the format easy to
write from within such a code.

PSDF objects are descriptions of \emph{data}, not instructions for
actions for an application.  In other words PSDF does not contain any
hidden ``language'' structures.  For example, there is no instruction
for ``adding'' or ``deleting'' a particle from the stream.  Adding
such instructions would require applications to implement interpreters
for implementing the instructions in the stream, which raises issues
of security and language design that would significantly complicate
the specification.

\subsection{Information-Rich}

The fundamental state of a point-particle can be specified in 8
numbers: one mass, one time, three position coordinates and three
velocity coordinates.  However, some auxiliary information about the
particle's state can be very helpful: accurate prediction of the
particle's position and velocity---for example, to display its track
in a visualization---can be facilitated by information about the
higher derivatives of its position in time.  When available to the
integration routine, these can be easily provided by our format (see
Table \ref{tab:names}).  

\subsection{Flexible}

Not all particles in many $N$-body simulations are point-masses!  For
example, simulations may attempt to model stellar evolution, and
therefore store ``star'' properties like radii and masses, or entropy
profiles with their particles.  Or, simulations may include fluid
particles subject to non-gravitational forces for SPH calculations.
To attempt to standardize names for every possible particle property
would result in a rigid and cumbersome format; instead, by allowing
arbitrary application-specific names in PSDF particle mappings that
can be ignored when not understood we permit complex applications to
work with application-specific data while ensuring that simple
applications can make use of the parts of the data they understand.

\subsection{Simple, Safe, and Composable}

PSDF documents are \emph{composable}, meaning that any two PSDF
streams can be concatenated to form another valid PSDF stream, and a
single stream can be split into a number of valid sub-streams.  It is
quite simple to write a short script in any number of languages that
consumes a PSDF stream and produces another recording the history of a
particular particle in the original stream, for example.  Another
example useful in practice is ``thinning'' a stream by including only
every $n$th particle update, when the interpolation requirements of
the consuming application are looser than those of the producing
application.  Adding header information, or instructions, or any other
meta-data to the stream defeats this goal by requiring specification
of some way to split and combine the associated meta-data.

\section{Repository}
\label{sec:repository}

We have examples of codes that generate and manipulate PSDF streams at 
\begin{verbatim}
https://github.com/jmakino/Particle-Stream-Data-Format
\end{verbatim}
The examples are in various different languages, and include
\begin{itemize}
\item Programs to generate initial conditions for $N$-body simulations
  in PSDF format.
\item Code and references to several different individual-time-step
  integrators that can take PSDF input and advance the corresponding
  system in time, producing PSDF output at each step.
\item Various post-processing tools that compute useful system
  properties from PSDF input.
\item A visualization program that takes PSDF input and produces a 3D
  representation of the system that can be played forward and
  backward, zoomed, etc.
\end{itemize}
We hope that the examples we provide will make it easy for the
community to use PSDF in their simulations, and that these users will,
in turn, contribute their useful programs back to the repository as
examples for future users.

\section*{Acknowledgments}
Part of the work was done while the authors visited the Center for
Planetary Science (CPS) in Kobe, Japan, during a visit that was funded
by the HPCI Strategic Program of MEXT.  We are grateful for their
hospitality.

\bibliographystyle{elsarticle-harv}
\bibliography{data-format}

\end{document}